# The Optimal Tradeoff Between PAPR and Ambiguity Functions for Generalized OFDM Waveform Set in ISAC Systems

Bichai Wang, Xiuhong Wei, Xueru Li, and Yongxing Zhou, *Fellow, IEEE*

*Abstract*—Integrated sensing and communications (ISAC) has been identified as one of the six usage scenarios for IMT-2030. Compared with communication performance, sensing performance is much more vulnerable to interference, and the received backscattered sensing signal with target information is usually too weak to be detected. It's interesting to understand the optimal tradeoff between interference rejection and signal strength improvement for the best sensing performance, but unfortunately it still remains unknown. In this paper, the trinity of auto-ambiguity function (AF), cross-AF and peak-to-average-power ratio (PAPR) is proposed to describe the interference and coverage related aspects for ISAC systems where multi-carrier waveform is usually assumed. We extend the existing orthogonal frequency division multiplexing (OFDM) waveforms in 5G to a generalized OFDM waveform set with some new members and a unified parametric representation. Then the optimal Pareto tradeoff between PAPR, auto-AF and cross-AF (i.e., the union bound) is developed for the generalized OFDM waveform set. To achieve the optimal Pareto union bound with reasonable computational complexity, we further propose a framework to optimize waveform parameters and sequences jointly. Finally, some practical design examples are provided and numerical results reveal that significant improvements can be achieved compared to the state-of-the-art 5G waveforms and sequences.

*Index Terms*—ISAC, waveform, sequence, PAPR, ambiguity function.

## I. INTRODUCTION

INTEGRATED sensing and communications (ISAC) has been identified by the radio communication division of the International Telecommunication Union (ITU-R) as one of the six usage scenarios for IMT-2030 [1][2]. Apart from communication capabilities, ISAC requires the support of sensing capabilities such as range estimation (i.e., delay estimation), velocity estimation (i.e., Doppler estimation), angle estimation and so on [3][4]. To promote ISAC standardization in 6G, the 3rd Generation Partnership Project (3GPP) has started the research on channel modeling for ISAC since Release 19 [5].

Performance of sensing is vulnerable to interference due to absence of effective protection mechanisms such as channel coding and decoding which have been commonly used in communication systems. In addition, the received sensing signal with target information (e.g., location, speed and trajectory, etc.) is usually too weak to be detected owing to small radar cross-section (RCS) and double propagation range in monostatic sensing applications. To achieve the optimal sensing performance, it is critical to understand the tradeoff between interference rejection and signal strength improvement.

From the signal strength point of view, a large peak-to-average-power ratio (PAPR) may degrade the maximum transmit power under practical power amplifiers (PAs). Hence, it is important to keep the PAPR of the transmitted signal low to ensure good sensing coverage, especially in the high-frequency band [6]. From the interference rejection point of view, ambiguity function (AF) is an important tool for analyzing the interference related aspects, which is a two-dimensional function of propagation delay and Doppler shift [7-10]. Note that AF instead of correlation function is considered here, because correlation function does not take Doppler shift into account and thus is not suitable for sensing scenarios with moving objects or targets. The impacts of the main lobe width of auto-AF, auto-AF peak sidelobe level (APSL), and cross-AF peak sidelobe level (CPSL) on sensing performance are summarized as follows:

- *Main lobe width of auto-AF:* For multi-target detection, the main lobe width of auto-AF reflects the capability to differentiate two or more targets with similar delay and Doppler parameters. The wider the main lobe of auto-AF, the harder it is to distinguish adjacent targets with almost the same velocity. In addition, a wide main lobe may result in ranging or velocity estimation errors, since the shape of the main lobe corresponding to the target is likely to be distorted by echo signals from objects near the target.
- *APSL:* APSL indicates how likely the main lobe corresponding to a target is to be overwhelmed by the sidelobes corresponding to other objects. Particularly, in monostatic sensing applications, when a target is farther away from the sensing transceiver than one or more other objects so that the echo signals from the other objects are stronger, a large APSL is likely to cause the main lobe corresponding to the target to be overwhelmed. This phenomenon is called the near-far effect, which has a strong negative impact on target detection accuracy. In addition, in bistatic sensing applications, the echo signal strength from a target may also be lower than that from other objects depending on the location of the target and the other objects. Therefore, APSL should be low enough to effectively suppress interference from objects with higher echo signal strength.

- *CPSL:* CPSL is capable of reflecting the ability to suppress interference from interfering transmitters. Due to limited time-frequency resources, it is inevitable that multiple transmitters transmit signals on the same time-frequency resources. When the interfering signal strength is higher than the echo signal strength from a target, a large CPSL may cause the main lobe corresponding to the target to be overwhelmed by the sidelobes corresponding to the interfering transmitters. It should be noted that the interfering signals are very likely to be stronger, especially in scenarios where there is a strong line-of-sight (LoS) path between the sensing receiver and the interfering transmitter. Therefore, a sufficient low CPSL is crucial to suppress interference from interfering transmitters.

Both PAPR and AF performance can be improved by carefully designing waveforms and sequences. Particularly, two OFDM waveforms, i.e., cyclic prefix orthogonal frequency division multiplexing (CP-OFDM) and discrete Fourier transform spread orthogonal frequency-division multiplexing (DFT-s-OFDM), are supported in 5G New Radio (NR) systems [11] [12]. Higher spectral efficiency can be obtained by CP-OFDM, while lower PAPR can be achieved by DFT-s-OFDM. To further reduce PAPR and thus improve communication coverage, frequency-domain spectral shaping (FDSS) can be implemented in 5G NR when using DFT-s-OFDM and $\pi/2$ binary phase shift keying (BPSK) modulation [13]. Apart from CP-OFDM and DFT-s-OFDM, inverse symplectic finite Fourier transform (ISFFT)-based orthogonal time-frequency space (OTFS) is another OFDM waveform that is resilient to the Doppler effect in high mobility scenarios [14]. In addition, a non-orthogonal waveform named faster-than-Nyquist (FTN) has been widely studied, and the potential of FTN to improve spectral efficiency and reduce PAPR has been demonstrated [15-17]. To analyze the sensing performance of communication-centric ISAC waveforms, a theoretical framework was established in [18] to evaluate the ranging performance under random data symbols, and the optimality of CP-OFDM waveform was proved. Nevertheless, only the auto-correlation function under integer delay was considered.

As for sequences, Zadoff-Chu (ZC) sequences and Gold sequences are widely adopted in 5G NR to generate reference signals for communications [12], where auto-correlation function and cross-correlation function under integer delay are usually used as metrics. However, the existing ZC and Gold sequences cannot provide a good interference rejection capability in practical ISAC systems where the Doppler effect should be considered. Active research in designing sensing sequences can also be found [7-10]. Particularly, several bounds have been developed to provide guidance for sequence design [9]. The Welch bound proposed in [19] was a lower bound on the larger value between the maximum auto-correlation magnitude and the maximum cross-correlation magnitude. Moreover, a tradeoff of the maximum auto-correlation magnitude and the maximum cross-correlation magnitude, i.e., the Sarwate bound, was proposed in [20]. However, both the Welch bound and the Sarwate bound only focused on correlations under integer delay, without considering the Doppler effect. Furthermore, by generalizing the Welch bound, a lower bound on the larger value between the maximum auto-AF magnitude and the maximum cross-AF magnitude under integer delay and integer Doppler shift was developed in [21]. Nevertheless, under the constraints of limited bandwidth and transmission duration, the assumption of integer delay and integer Doppler shift is obviously impractical in sensing scenarios. More importantly, the optimal tradeoff between PAPR and AFs is still unclear.

In this paper, the trinity of auto-AF, cross-AF and PAPR is proposed to describe the interference and coverage related aspects for ISAC systems, and the optimal tradeoff between PAPR and AFs under fractional delay and fractional Doppler shift is investigated. The main contributions of this paper can be summarized as follows:

- We extend the existing CP-OFDM and DFT-s-OFDM waveforms to a generalized OFDM waveform set with some new members and a unified parametric representation. Specifically, a preprocessing operation based on several preprocessing matrices is performed on information symbols (e.g., sequences or modulated data symbols) before subcarrier mapping, and different preprocessing matrices may be adopted to generate different waveforms. The waveform parameters, including the preprocessing matrices, and sequences can be further jointly designed to achieve the best sensing performance.
- The optimal Pareto tradeoff between PAPR, auto-AF and cross-AF (i.e., the union bound) under fractional delay and fractional Doppler shift is developed for the generalized OFDM waveform set, which can provide guidance for the design of waveform parameters and sequences. It's also shown that the performance of the existing 5G waveforms and sequences still lags far behind the optimal Pareto union bound. Therefore, it is meaningful to explore better schemes.
- To achieve the optimal Pareto union bound on PAPR, auto-AF and cross-AF with reasonable computational complexity, we further propose a framework to optimize waveform parameters and sequences jointly. Specifically, under fractional delay and fractional Doppler shift constraints, a multi-objective optimization problem is formulated to achieve a tradeoff between PAPR, APSL and CPSL. Then, an

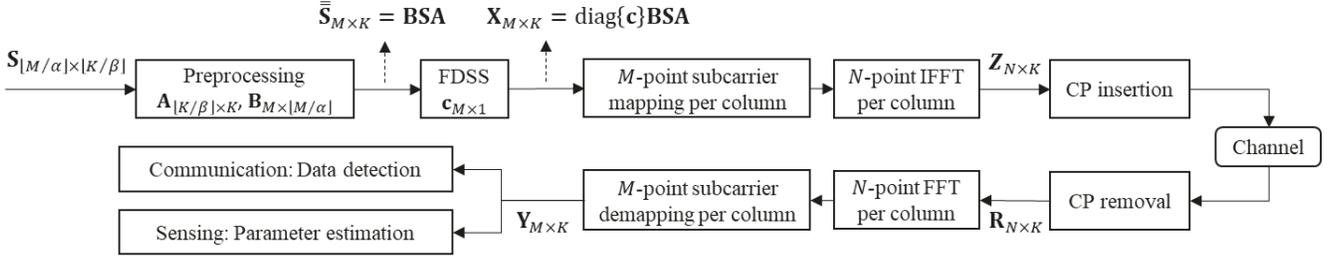

**Fig. 1.** Transceiver structure of the generalized OFDM waveform set.

TABLE I
SPECIAL CASES IN THE GENERALIZED OFDM WAVEFORM SET

| Waveform | $\alpha$ | $\beta$ | $\mathbf{A}_{\lfloor K/\beta \rfloor \times K}$ | $\mathbf{B}_{M \times \lfloor M/\alpha \rfloor}$ | $\mathbf{c}_{M \times 1}$ |
|---|---|---|---|---|---|
| CP-OFDM | $\alpha = 1$ | $\beta = 1$ | $\mathbf{I}_K$ | $\mathbf{I}_M$ | Optimizable |
| DFT-s-OFDM | $\alpha = 1$ | $\beta = 1$ | $\mathbf{I}_K$ | $\mathbf{F}_M$ | Optimizable |
| OTFS | $\alpha = 1$ | $\beta = 1$ | $\mathbf{F}_K^H$ | $\mathbf{F}_M$ | Optimizable |
| FTN-s-OFDM | $0 < \alpha < 1$ | $\beta = 1$ | $\mathbf{I}_K$ | $\mathbf{F}_{\lfloor M/\alpha \rfloor}(\mathcal{M},:)$ | Optimizable |
| FTN-s-OTFS | $0 < \alpha < 1$ | $\beta = 1$ | $\mathbf{F}_K^H$ | $\mathbf{F}_{\lfloor M/\alpha \rfloor}(\mathcal{M},:)$ | Optimizable |
| DFTN-s-OTFS | $0 < \alpha < 1$ | $0 < \beta < 1$ | $\mathbf{F}_{\lfloor K/\beta \rfloor}^H(:,\mathcal{K})$ | $\mathbf{F}_{\lfloor M/\alpha \rfloor}(\mathcal{M},:)$ | Optimizable |

- adaptive waveform and sequence generation network (AWSGNet) is proposed to solve this multi-objective problem.
- Some practical design examples taking realistic engineering aspects into account are provided, where different types of sequences are considered, e.g., sequences with continuous amplitude and continuous phase, unimodular sequences with continuous phase, as well as unimodular sequences with discrete phase. The numerical results reveal that significant improvements can be achieved compared to the state-of-the-art 5G waveforms and sequences.

The remainder of the paper is structured as follows. Section II introduces the generalized OFDM waveform set. Then, the optimal Pareto union bound on PAPR, auto-AF and cross-AF is described in Section III. Section IV presents the proposed framework for optimizing waveform parameters and sequences, and some design examples are provided in Section V. Finally, the conclusions are drawn in Section VI.

## II. GENERALIZED OFDM WAVEFORM SET

In ISAC systems, it is expected that different waveforms can be flexibly configured in different scenarios. For example, CP-OFDM can be used in high-throughput scenarios, while DFT-s-OFDM is preferred when coverage is more important. In addition, it is desirable to introduce new waveforms to effectively balance communication and sensing performance. To this end, we extend the existing CP-OFDM and DFT-s-OFDM waveforms to a generalized OFDM waveform set with some new members and a unified parametric representation. The transceiver structure of the generalized OFDM waveform set is shown in Fig. 1.

In the transmitter, $\lfloor M/\alpha \rfloor \times \lfloor K/\beta \rfloor$ information symbols, denoted as $\mathbf{S}_{\lfloor M/\alpha \rfloor \times \lfloor K/\beta \rfloor}$, are generated, where $0 < \alpha \leq 1$, $0 < \beta \leq 1$, $M$ is the number of subcarriers and $K$ is the number of OFDM symbols allocated for transmission. After performing the preprocessing operation, i.e., a plurality of matrix multiplication operations, the time-frequency domain signal $\mathbf{X}_{M \times K} = [\mathbf{x}_0, \mathbf{x}_1, \cdots, \mathbf{x}_{K-1}]$ can be obtained by

$$\mathbf{X}_{M \times K} = \text{diag}\{\mathbf{c}_{M \times 1}\} \mathbf{B}_{M \times \lfloor M/\alpha \rfloor} \mathbf{S}_{\lfloor M/\alpha \rfloor \times \lfloor K/\beta \rfloor} \mathbf{A}_{\lfloor K/\beta \rfloor \times K}, \quad (1)$$

where $\mathbf{c}_{M \times 1}$ is the FDSS vector. Particularly, if FDSS is not applied, we have $\text{diag}\{\mathbf{c}_{M \times 1}\} = \mathbf{I}_M$, where $\mathbf{I}_M$ represents identity matrix of size $M \times M$. The preprocessing matrices $\mathbf{A}_{\lfloor K/\beta \rfloor \times K}$ and $\mathbf{B}_{M \times \lfloor M/\alpha \rfloor}$, the FDSS vector $\mathbf{c}_{M \times 1}$, as well as the signal $\mathbf{S}_{\lfloor M/\alpha \rfloor \times \lfloor K/\beta \rfloor}$ can be jointly designed to achieve a tradeoff between PAPR and AFs, as detailed in Section III and Section IV. After subcarrier mapping, $N$-point inverse fast Fourier transformation (IFFT) is performed per OFDM symbol to obtain the delay-time domain signal $\mathbf{Z}_{N \times K} = [\mathbf{z}_0, \mathbf{z}_1, \cdots, \mathbf{z}_{K-1}]$. Specifically, the $n$th element of $\mathbf{z}_k = [z_{0,k}, z_{1,k}, \cdots, z_{N-1,k}]^T$ can be expressed as

$$z_{n,k} = \frac{1}{\sqrt{N}} \sum_{i \in \aleph} x_{q_i, k} e^{j \frac{2 \pi i n}{N}}, \quad (2)$$

where $n \in \{0, 1, \cdots, N-1\}$, $k \in \{0, 1, \cdots, K-1\}$, $q_i \in \{0, 1, \cdots, M-1\}$, $x_{q_i, k}$ is the $q_i$th element of $\mathbf{x}_k = [x_{0,k}, x_{1,k}, \cdots, x_{M-1,k}]^T$, $\aleph$ is the subcarrier index set corresponding to the allocated $M$ subcarriers, and the subcarrier index of the $q_i$th subcarrier in the $M$ subcarriers is $i$. Without loss of generality, it is assumed that $\aleph = \{0, 1, \cdots, M-1\}$ in this paper, and thus we have $q_i = i$. Then, a cyclic prefix (CP) is inserted before each OFDM symbol to avoid inter-symbol interference (ISI).

In the receiver, the received delay-time domain signal $\mathbf{R}_{N \times K} = [\mathbf{r}_0, \mathbf{r}_1, \cdots, \mathbf{r}_{K-1}]$ can be obtained after CP removal, and the received signal $\mathbf{r}_k$ on the $k$th OFDM symbol can be expressed as

$$\mathbf{r}_k = \mathbf{h}_k * \mathbf{z}_k + \mathbf{v}_k, \quad (3)$$

where $\mathbf{h}_k$ is the channel impulse response vector on the $k$th OFDM symbol, $\mathbf{v}_k$ is the noise vector on the $k$th OFDM symbol, and "$*$" represents the cyclic convolution operation. By performing $N$-point fast Fourier transformation (FFT) and subcarrier demapping per OFDM symbol, the received time-frequency domain signal $\mathbf{Y}_{M \times K} = [\mathbf{y}_0, \mathbf{y}_1, \cdots, \mathbf{y}_{K-1}]$ can be obtained. Then, data detection or sensing parameter estimation can be performed based on $\mathbf{Y}_{M \times K}$. For example, the conventional periodogram-based estimation algorithm [22] can be used for sensing based on the received signal $\mathbf{Y}_{M \times K}$ and the transmitted signal $\mathbf{X}_{M \times K}$, which is also equivalent to the result based on the matched-filtering output between the transmitted delay-time domain signal and the corresponding echo signal. Particularly, if FDSS is applied at the transmitter and the FDSS vector is known at the receiver, the received time-frequency domain signal $\mathbf{Y}_{M \times K}$ can be processed based on the FDSS vector, i.e., $\widetilde{\mathbf{Y}}_{M \times K} = \text{diag}\{(\mathbf{c}_{M \times 1})^*\}\mathbf{Y}_{M \times K}$, where $()^*$ denotes the conjugate operation. Then, $\widetilde{\mathbf{Y}}_{M \times K}$ can be used for data detection or sensing parameter estimation.

As shown in Table I, the existing CP-OFDM and DFT-s-OFDM waveforms in 5G communication systems belong to the generalized OFDM waveform set. In addition, the generalized OFDM waveform set includes OTFS and several new FTN-based waveforms such as FTN spread OFDM (FTN-s-OFDM), FTN spread OTFS (FTN-s-OTFS) and double FTN spread OTFS (DFTN-s-OTFS).

***Remark 1:*** In 5G communication systems, FDSS can be used in DFT-s-OFDM waveform to reduce PAPR, especially when $\pi/2$ BPSK modulation is adopted to generate communication symbols. However, it is not sufficient to only consider the PAPR performance in ISAC systems. A well-designed FDSS vector may have a positive impact on both PAPR and AFs. Some performance comparisons with and without FDSS will be provided in Section V.

*A. CP-OFDM*

In CP-OFDM waveform, we have $\alpha = 1$, $\beta = 1$, $\mathbf{A}_{\lfloor K/\beta \rfloor \times K} = \mathbf{I}_K$ and $\mathbf{B}_{M \times \lfloor M/\alpha \rfloor} = \mathbf{I}_M$. Particularly, the auto-AF is a two-dimensional sinc-like function if $\text{diag}\{\mathbf{c}_{M \times 1}\} = \mathbf{I}_M$ and unimodular sequences are generated in the time-frequency domain, which means that the auto-AF shape cannot be changed by only optimizing the sequence elements. Therefore, it is significant to design FDSS to improve sensing performance.

*B. DFT-s-OFDM*

In DFT-s-OFDM waveform, $\mathbf{S}_{\lfloor M/\alpha \rfloor \times \lfloor K/\beta \rfloor}$ is generated in the delay-time domain, and DFT is performed to obtain the time-frequency domain signal $\mathbf{X}_{M \times K}$. In this case, we have $\alpha = 1$, $\beta = 1$, $\mathbf{A}_{\lfloor K/\beta \rfloor \times K} = \mathbf{I}_K$ and $\mathbf{B}_{M \times \lfloor M/\alpha \rfloor} = \mathbf{F}_M$, where $\mathbf{F}_M$ represents DFT matrix of size $M \times M$, and the element in the $m$th row and the $i$th column of $\mathbf{F}_M$ is $1/\sqrt{M}\, e^{-j2\pi im/M}$.

*C. OTFS*

OTFS can be implemented by performing ISFFT on the basis of CP-OFDM, where $\mathbf{S}_{\lfloor M/\alpha \rfloor \times \lfloor K/\beta \rfloor}$ is generated in the delay-Doppler domain. Specifically, in OTFS waveform, we have $\alpha = 1$, $\beta = 1$, $\mathbf{A}_{\lfloor K/\beta \rfloor \times K} = \mathbf{F}_K^H$ and $\mathbf{B}_{M \times \lfloor M/\alpha \rfloor} = \mathbf{F}_M$, where $\mathbf{F}_K$ and $\mathbf{F}_M$ represent DFT matrices of size $K \times K$ and $M \times M$, respectively.

*D. FTN-s-OFDM*

FTN has been widely studied to improve spectral efficiency and reduce PAPR in communication scenarios. To be compatible with existing OFDM systems and maintain the advantages of FTN, FTN-s-OFDM waveform is developed.

In FTN-s-OFDM waveform, we have $0 < \alpha < 1$, $\beta = 1$, $\mathbf{A}_{\lfloor K/\beta \rfloor \times K} = \mathbf{I}_K$ and $\mathbf{B}_{M \times \lfloor M/\alpha \rfloor} = \mathbf{F}_{\lfloor M/\alpha \rfloor}(\mathcal{M},:)$, where $\mathbf{F}_{\lfloor M/\alpha \rfloor}$ represents DFT matrix of size $\lfloor M/\alpha \rfloor \times \lfloor M/\alpha \rfloor$. $\mathbf{F}_{\lfloor M/\alpha \rfloor}(\mathcal{M},:)$ is a matrix consisting of $M$ different rows of $\mathbf{F}_{\lfloor M/\alpha \rfloor}$, and $\mathcal{M}$ is the row index set corresponding to these $M$ different rows where $\mathcal{M} \subseteq \{0, 1, \cdots, \lfloor M/\alpha \rfloor - 1\}$. Without loss of generality, we assume that $\mathcal{M} = \{0, 1, \cdots, M - 1\}$, then the $m$th element of $\mathbf{x}_k$ can be expressed as

$$x_{m,k} = c_m \cdot \frac{1}{\sqrt{\lfloor M/\alpha \rfloor}} \sum_{i=0}^{\lfloor M/\alpha \rfloor - 1} s_{i,k} e^{-j\frac{2\pi i m}{\lfloor M/\alpha \rfloor}}, \quad (4)$$

where $m \in \{0, 1, \cdots, M - 1\}$, $c_m$ is the $m$th element of $\mathbf{c}_{M \times 1} = [c_0, c_1, \cdots, c_{M-1}]^T$, and $s_{i,k}$ is the $i$th element of $\mathbf{s}_k = [s_{0,k}, s_{1,k}, \cdots, s_{\lfloor M/\alpha \rfloor - 1, k}]^T$, where $\mathbf{s}_k$ is the $k$th column of the delay-time domain signal $\mathbf{S}_{\lfloor M/\alpha \rfloor \times \lfloor K/\beta \rfloor}$, i.e., $\mathbf{S}_{\lfloor M/\alpha \rfloor \times \lfloor K/\beta \rfloor} = [\mathbf{s}_0, \mathbf{s}_1, \cdots, \mathbf{s}_{\lfloor K/\beta \rfloor - 1}]$. Then, by substituting (4) into (2), $z_{n,k}$ can be rewritten as

$$\begin{aligned} z_{n,k} &= \frac{1}{\sqrt{N}} \sum_{t=0}^{M-1} \frac{1}{\sqrt{\lfloor M/\alpha \rfloor}} \sum_{i=0}^{\lfloor M/\alpha \rfloor - 1} c_t s_{i,k} e^{-j\frac{2\pi it}{\lfloor M/\alpha \rfloor}} e^{j\frac{2\pi tn}{N}} \\ &= \frac{1}{\sqrt{\lfloor M/\alpha \rfloor}} \sum_{i=0}^{\lfloor M/\alpha \rfloor - 1} s_{i,k} \frac{1}{\sqrt{N}} \sum_{t=0}^{M-1} c_t e^{j\frac{2\pi t}{N}\left(n - \frac{iN}{\lfloor M/\alpha \rfloor}\right)} \\ &= \frac{1}{\sqrt{\lfloor M/\alpha \rfloor}} \sum_{i=0}^{\lfloor M/\alpha \rfloor - 1} s_{i,k} g\left(n - \frac{iN}{\lfloor M/\alpha \rfloor}\right), \end{aligned} \quad (5)$$

where $g(u)$ is the effective pulse-shape function. Particularly, if FDSS is not applied, $g(u)$ can be expressed as

$$g(u) = \frac{1}{\sqrt{N}} \sum_{t=0}^{M-1} e^{j\frac{2\pi tu}{N}} = \frac{1}{\sqrt{N}} e^{j\frac{\pi(M-1)u}{N}} \frac{\sin\left(\frac{\pi M u}{N}\right)}{\sin\left(\frac{\pi u}{N}\right)}. \quad (6)$$

It can be seen from (5) and (6) that the effective pulse-shape function $g(u)$ is a sinc-like function with the first null at $N/M$, and is modulated by $\lfloor M/\alpha \rfloor$ complex symbols per OFDM symbol with the corresponding time shifts, i.e., $iN/\lfloor M/\alpha \rfloor$ for the $i$th complex symbol. Therefore, FTN-s-OFDM is a special implementation of the time-domain FTN if $0 < \alpha < 1$, where $\alpha$ can be regarded as the FTN time-domain compression factor [15]. Note that FTN-s-OFDM degenerates into DFT-s-OFDM waveform if $\alpha = 1$. Since $\mathbf{F}_{\lfloor M/\alpha \rfloor}(\mathcal{M},:)\mathbf{s}_k$ is equivalent to truncating the $\lfloor M/\alpha \rfloor$-length signal after DFT to the $M$-length signal in the frequency domain, $\alpha$ can also be referred to as frequency-domain truncation factor. By adjusting the value of $\alpha$, the time shifts can be changed accordingly. As a result, PAPR can be reduced by selecting an appropriate value of $\alpha$.

***Remark 2:*** In communication scenarios, a small value of $\alpha$ may result in poor bit error performance. For example, a significant signal-to-noise ratio (SNR) loss can be observed at a specific block error rate (BLER) performance when $\alpha < 0.8$ [15]. However, in sensing scenarios, a small value of $\alpha$ can be conducive to achieving good PAPR and AF performance, which will be verified in Section V. Therefore, different values of $\alpha$ should be selected for communication and sensing scenarios.

*E. FTN-s-OTFS*

Similar to FTN-s-OFDM, we propose to integrate FTN with OTFS waveform, which is referred to as FTN-s-OTFS waveform in this paper. More specifically, let $0 < \alpha < 1$, $\beta = 1$, $\mathbf{A}_{\lfloor K/\beta \rfloor \times K} = \mathbf{F}_K^H$ and $\mathbf{B}_{M \times \lfloor M/\alpha \rfloor} = \mathbf{F}_{\lfloor M/\alpha \rfloor}(\mathcal{M},:)$, then we have $\mathbf{X}_{M \times K} = \text{diag}\{\mathbf{c}_{M \times 1}\} \mathbf{F}_{\lfloor M/\alpha \rfloor}(\mathcal{M},:) \mathbf{S}_{\lfloor M/\alpha \rfloor \times \lfloor K/\beta \rfloor} \mathbf{F}_K^H$, where $\mathbf{S}_{\lfloor M/\alpha \rfloor \times \lfloor K/\beta \rfloor}$ is generated in the delay-Doppler domain and frequency-domain truncation is performed after ISFFT.

*F. DFTN-s-OTFS*

Inspired by FTN-s-OTFS with frequency-domain truncation, DFTN-s-OTFS waveform is developed by performing both frequency-domain truncation and Doppler-domain truncation, where $\beta$ can be referred to as Doppler-domain truncation factor. In DFTN-s-OTFS waveform, we have $0 < \alpha < 1$, $0 < \beta < 1$, $\mathbf{A}_{\lfloor K/\beta \rfloor \times K} = \mathbf{F}_{\lfloor K/\beta \rfloor}^H(:,\mathcal{K})$ and $\mathbf{B}_{M \times \lfloor M/\alpha \rfloor} = \mathbf{F}_{\lfloor M/\alpha \rfloor}(\mathcal{M},:)$, where $\mathbf{F}_{\lfloor K/\beta \rfloor}$ represents DFT matrix of size $\lfloor K/\beta \rfloor \times \lfloor K/\beta \rfloor$. $\mathbf{F}_{\lfloor K/\beta \rfloor}^H(:,\mathcal{K})$ is a matrix consisting of $K$ different columns of $\mathbf{F}_{\lfloor K/\beta \rfloor}^H$, and $\mathcal{K}$ is the column index set corresponding to these $K$ different columns where $\mathcal{K} \subseteq \{0,1,\cdots,\lfloor K/\beta \rfloor - 1\}$. Without loss of generality, we assume that the first $K$ columns of $\mathbf{F}_{\lfloor K/\beta \rfloor}^H$ are selected, then we have $\mathcal{K} = \{0,1,\cdots, K-1\}$.

## III. THE OPTIMAL PARETO UNION BOUND ON PAPR AND AFS FOR THE GENERALIZED OFDM WAVEFORM SET

As mentioned before, low PAPR and good AFs are critical to superior sensing performance. However, it is very challenging to achieve the optimal PAPR, auto-AF and cross-AF performance simultaneously. The best tradeoff between these metrics is still an open issue, especially under practical fractional delay and fractional Doppler shift constraints. To fill this gap, the optimal Pareto union bound on PAPR, auto-AF and cross-AF under fractional delay and fractional Doppler shift is developed for the generalized OFDM waveform set, which can provide guidance for the design of waveform parameters and sequences in ISAC systems.

Let $\mathbf{U}$ be a sequence group set consisting of $D$ sequence groups, where each sequence group consists of $\lfloor K/\beta \rfloor$ sequences of length $\lfloor M/\alpha \rfloor$. A sequence matrix $\mathbf{S}_{\lfloor M/\alpha \rfloor \times \lfloor K/\beta \rfloor}^{\mathbf{P}}$ can be generated according to the sequence group $\mathbf{P}$ in $\mathbf{U}$, and different columns in $\mathbf{S}_{\lfloor M/\alpha \rfloor \times \lfloor K/\beta \rfloor}^{\mathbf{P}}$ correspond to different sequences in $\mathbf{P}$. Based on the generalized OFDM waveform set proposed in Section II, by replacing $\mathbf{S}_{\lfloor M/\alpha \rfloor \times \lfloor K/\beta \rfloor}$ with $\mathbf{S}_{\lfloor M/\alpha \rfloor \times \lfloor K/\beta \rfloor}^{\mathbf{P}}$, the time-frequency domain signal $\mathbf{X}_{M \times K}^{\mathbf{P}} = [\mathbf{x}_0^{\mathbf{P}}, \mathbf{x}_1^{\mathbf{P}}, \cdots, \mathbf{x}_{K-1}^{\mathbf{P}}]$ and the delay-time domain signal $\mathbf{Z}_{N \times K}^{\mathbf{P}} = [\mathbf{z}_0^{\mathbf{P}}, \mathbf{z}_1^{\mathbf{P}}, \cdots, \mathbf{z}_{K-1}^{\mathbf{P}}]$ can be obtained according to (1) and (2), respectively. Without loss of generality, we assume that $\mathbf{Z}_{N \times K}^{\mathbf{P}}$ is transmitted on $K$ different OFDM symbols with the same time interval, which is denoted as $T_c$.

*Definition 1:* The maximum PAPR corresponding to sequence group set $\mathbf{U}$ is defined as

$$PAPR_U = \max_{\forall \mathbf{P} \in U, \forall k} PAPR_{\mathbf{z}_k^{\mathbf{P}}}, \quad (7)$$

where the PAPR of $\mathbf{z}_k^{\mathbf{P}} = [z_{0,k}^{\mathbf{P}}, z_{1,k}^{\mathbf{P}}, \cdots, z_{N-1,k}^{\mathbf{P}}]^T$ can be calculated by

$$PAPR_{\mathbf{z}_k^{\mathbf{P}}} = \frac{\max_{n=0,1,\cdots,N-1} |z_{n,k}^{\mathbf{P}}|^2}{\frac{1}{N} \sum_{i=0}^{N-1} |z_{i,k}^{\mathbf{P}}|^2}. \quad (8)$$

Since CP is inserted before each OFDM symbol, the periodic AF is considered in this paper. Similar to the generation of $\mathbf{X}_{M \times K}^{\mathbf{P}}$ and $\mathbf{Z}_{N \times K}^{\mathbf{P}}$ corresponding to sequence group $\mathbf{P}$, the time-frequency domain signal $\mathbf{X}_{M \times K}^{\mathbf{Q}} = [\mathbf{x}_0^{\mathbf{Q}}, \mathbf{x}_1^{\mathbf{Q}}, \cdots, \mathbf{x}_{K-1}^{\mathbf{Q}}]$ and the delay-time domain signal $\mathbf{Z}_{N \times K}^{\mathbf{Q}} = [\mathbf{z}_0^{\mathbf{Q}}, \mathbf{z}_1^{\mathbf{Q}}, \cdots, \mathbf{z}_{K-1}^{\mathbf{Q}}]$ corresponding to sequence group $\mathbf{Q}$ in $\mathbf{U}$ can also be obtained.

*Definition 2:* The periodic AF of sequence groups $\mathbf{P}$ and $\mathbf{Q}$ in $\mathbf{U}$ is defined as

$$AF_{\mathbf{P},\mathbf{Q}}^U(\tau, f_d) = \sum_{k=0}^{K-1} AF_{\mathbf{z}_k^{\mathbf{P}}, \mathbf{z}_k^{\mathbf{Q}}}(\tau, f_d) e^{j2\pi f_d k T_c}, \quad (9)$$

where $\tau \in \{0,1,2,\cdots,N-1\}$ is the delay index, $f_d \in [-f_D, f_D]$ is the Doppler shift and $f_D$ is the maximum Doppler shift considered. The delay interval $T_s$ and the Doppler shift interval $\Delta f_{ds}$ can be represented by $T_s = 1/(N\Delta f)$ and $\Delta f_{ds} = 2f_D/(J-1)$, respectively, where $\Delta f$ is the subcarrier spacing (SCS) and $J$ is the number of Doppler shifts. Furthermore, the periodic AF of $\mathbf{z}_k^{\mathbf{P}}$ and $\mathbf{z}_k^{\mathbf{Q}}$, i.e., $AF_{\mathbf{z}_k^{\mathbf{P}}, \mathbf{z}_k^{\mathbf{Q}}}(\tau, f_d)$, can be expressed as

$$AF_{\mathbf{z}_k^{\mathbf{P}}, \mathbf{z}_k^{\mathbf{Q}}}(\tau, f_d) = \sum_{n=0}^{N-1} z_{n,k}^{\mathbf{P}} (z_{(n+\tau) \bmod(N), k}^{\mathbf{Q}})^* e^{j2\pi f_d n T_s}, \quad (10)$$

where $\bmod()$ represents the modulo operation. Particularly, $AF_{\mathbf{P},\mathbf{Q}}^U(\tau, f_d)$ is called auto-AF if $\mathbf{P} = \mathbf{Q}$, otherwise $AF_{\mathbf{P},\mathbf{Q}}^U(\tau, f_d)$ is called cross-AF.

Note that we assume $T_s < 1/(M\Delta f)$ and $\Delta f_{ds} < 1/(KT_c)$ in this paper, i.e., the delay interval is less than the delay resolution and the Doppler shift interval is less than the Doppler resolution, which means fractional delay and fractional Doppler shift are considered. Based on the definition of AF in (9), the main lobe width of auto-AF, APSL, and CPSL under fractional delay and fractional Doppler shift can be further defined.

*Definition 3:* The main lobe width of auto-AF along the delay dimension (i.e., $W_1$) and the main lobe width of auto-AF along the Doppler dimension (i.e., $W_2$) can be defined as

$$W_1 = \frac{bM}{N},$$
$$W_2 = f_b K T_c, \quad (11)$$

where $b \in (0, \lfloor N/2 \rfloor]$ is the delay index corresponding to the first local minimum auto-AF magnitude when $f_d = 0$, and $f_b \in (0, f_D]$ is the Doppler shift corresponding to the first local minimum auto-AF magnitude when $\tau = 0$. Note that the same transmit power is usually assumed on the $K$ OFDM symbols. In this case, no matter what waveforms and sequences are used, the auto-AF along the Doppler dimension is a sinc-like function if $\tau = 0$ and thus we have $W_2 \approx 1$. Therefore, only the main lobe width of auto-AF along the delay dimension will be discussed below.

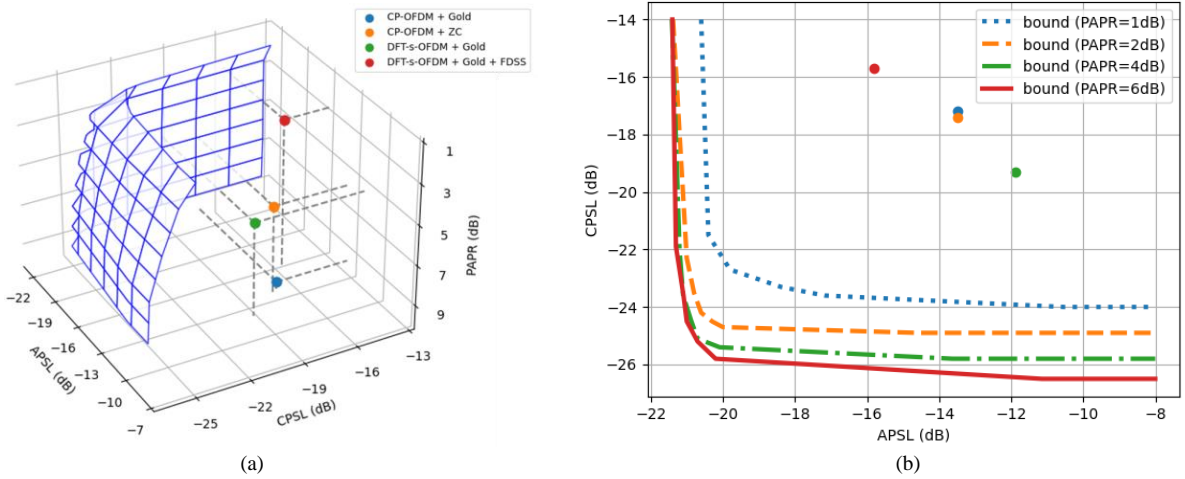

**Fig. 2.** The optimal Pareto union bound on PAPR, APSL, and CPSL: (a) The 3D surface of PAPR, APSL, and CPSL; (b) The 2D projection of the 3D surface on the APSL-CPSL plane.

***Definition 4:*** The APSL corresponding to sequence group set $U$ is defined as

$$APSL_U = \max_{\mathbf{P} \in U} \max_{\substack{\tau \notin [0,b] \cup [N-b, N-1] \\ f_d \notin [-f_b, f_b]}} \left| AF_{\mathbf{P},\mathbf{P}}^{U}(\tau, f_d) \right|. \quad (12)$$

***Definition 5:*** The CPSL corresponding to sequence group set $U$ is defined as

$$CPSL_U = \max_{\substack{\mathbf{P},\mathbf{Q} \in U \\ \mathbf{P} \neq \mathbf{Q}}} \max_{\substack{\forall \tau \\ f_d \in [-f_D, f_D]}} \left| AF_{\mathbf{P},\mathbf{Q}}^{U}(\tau, f_d) \right|. \quad (13)$$

We can see from (7) and (9) that the performance of PAPR and AFs directly depends on the delay-time domain signal $\mathbf{Z}_{N \times K}^{\mathbf{P}}, \forall \mathbf{P} \in U$. Since the time-frequency domain signal $\mathbf{X}_{M \times K}^{\mathbf{P}}$ and the delay-time domain signal $\mathbf{Z}_{N \times K}^{\mathbf{P}}$ can be converted to each other, the optimal Pareto union bound on PAPR and AFs can be found by traversing all possible $\mathbf{X}_{M \times K}^{\mathbf{P}}, \forall \mathbf{P} \in U$.

Without any constraints on the amplitude and phase values of the elements in $\mathbf{X}_{M \times K}^{\mathbf{P}}$, an example of the optimal Pareto union bound on PAPR, APSL and CPSL is shown in Fig. 2, where $D = 30$, $K = 4$, $M = 204$, $N = 2048$ and $W_1 \approx 1$. In addition, the performance of several existing 5G NR waveforms and sequences is also shown in Fig. 2 for comparison. Specifically, "CP-OFDM + Gold" and "CP-OFDM + ZC" represent the performance of CP-OFDM waveform using Gold and ZC sequences, respectively. "DFT-s-OFDM + Gold" and "DFT-s-OFDM + Gold + FDSS" indicate the performance of DFT-s-OFDM waveform with and without FDSS, respectively, where Gold sequences with $\pi/2$ BPSK modulation are adopted and the FDSS vector is generated based on the Root Raised Cosine (RRC) filter [13]. In the "CP-OFDM + Gold", "CP-OFDM + ZC" and "DFT-s-OFDM + Gold" schemes, we have $W_1 \approx 1$, while $W_1 \approx 1.4$ in the "DFT-s-OFDM + Gold + FDSS" scheme. Some insights are highlighted as follows:

- The performance of the existing 5G NR waveforms and sequences still lags far behind the optimal Pareto union bound on PAPR, APSL and CPSL. As shown in Fig. 2, although low PAPR can be achieved by the "DFT-s-OFDM + Gold + FDSS" scheme at the cost of increasing the main lobe width of auto-AF, the performance of APSL and CPSL under this scheme is still far from the optimal bound.

- The inherent tradeoff between PAPR, auto-AF and cross-AF is clearly shown in Fig. 2. Specifically, the optimal values of auto-AF and cross-AF vary with different PAPR constraints. For example, when the maximum PAPR is suppressed from 4 dB to 1 dB, the optimal performance of auto-AF and cross-AF is attenuated by about 0.8 dB and 1.8 dB, respectively. In addition, there is a tradeoff between auto-AF and cross-AF under a specific PAPR constraint, i.e., APSL increases with the decrease of CPSL, and vice versa.

- Nevertheless, it is quite surprising that the APSL-CPSL curve has an "L"-shape as shown in Fig. 2 (b). That is to say, an aggressive decrease in CPSL (or APSL) will only result in a marginal increase in APSL (or CPSL) until a turning point is reached. However, after the turning point, a small decrease in CPSL (or APSL) may lead to a catastrophic deterioration in APSL (or CPSL). Note that the turning point is defined as the point where the slope of the APSL-CPSL curve in Fig. 2 (b) is equal to -1. The value of APSL (or CPSL) corresponding to the turning point is close to the optimal value of APSL (or CPSL), which means a good tradeoff between APSL and CPSL can be achieved. Besides, the reduction of PAPR does not have a significant negative impact on the optimal performance of APSL and CPSL when the PAPR constraint is not too harsh. Therefore, in scenarios where PAPR, APSL and CPSL are all important, it looks promising to optimize APSL and CPSL under a low but not-too-extreme PAPR.

An interesting and important question is what kind of waveforms and sequences can achieve the optimal Pareto union bound. The following Theorem 1 is proposed to reveal the reachability of the optimal Pareto union bound on PAPR and AFs.

***Theorem 1***: Given a waveform with preprocessing matrices $\mathbf{A}_{\lfloor K/\beta \rfloor \times K}$ and $\mathbf{B}_{M \times \lfloor M/\alpha \rfloor}$, as well as FDSS vector $\mathbf{c}_{M \times 1}$, an optimal sequence group set $\mathbf{U}^{opt}$ can always be found to achieve the optimal Pareto union bound on PAPR and AFs, if the following condition is satisfied:

$$rank(\mathbf{A}_{\lfloor K/\beta \rfloor \times K}) = K,$$
$$rank(\text{diag}\{\mathbf{c}_{M \times 1}\}\mathbf{B}_{M \times \lfloor M/\alpha \rfloor}) = M, \quad (14)$$

where $rank()$ represents the rank of a matrix. The condition in (14) indicates that $\mathbf{A}_{\lfloor K/\beta \rfloor \times K}$ is a full-column-rank matrix and $\text{diag}\{\mathbf{c}_{M \times 1}\}\mathbf{B}_{M \times \lfloor M/\alpha \rfloor}$ is a full-row-rank matrix. Particularly, if FDSS is not applied, i.e., $\text{diag}\{\mathbf{c}_{M \times 1}\} = \mathbf{I}_M$, $\mathbf{B}_{M \times \lfloor M/\alpha \rfloor}$ should be a full-row-rank matrix.

***Proof***: It's assumed that the optimal Pareto union bound is achieved under the time-frequency domain signal set $\{\mathbf{X}_{M \times K}^d\}_{\forall d \in \mathcal{D}}$, where $\mathcal{D} = \{0, 1, \cdots, D\}$. If the condition in (14) is satisfied, the sequence matrix $\mathbf{S}_{\lfloor M/\alpha \rfloor \times \lfloor K/\beta \rfloor}^d$ corresponding to the $d$th sequence group in $\mathbf{U}^{opt}$ can be obtained by

$$\mathbf{S}_{\lfloor M/\alpha \rfloor \times \lfloor K/\beta \rfloor}^d = \mathbf{G}_r \mathbf{X}_{M \times K}^d \mathbf{G}_l, \quad (15)$$

where

$$\mathbf{G}_l = (\mathbf{A}_{\lfloor K/\beta \rfloor \times K}^H \mathbf{A}_{\lfloor K/\beta \rfloor \times K})^{-1} \mathbf{A}_{\lfloor K/\beta \rfloor \times K}^H, \quad (16)$$

and

$$\mathbf{G}_r = \left(\text{diag}\{\mathbf{c}_{M \times 1}\}\mathbf{B}_{M \times \lfloor M/\alpha \rfloor}\right)^H \cdot$$
$$\left(\text{diag}\{\mathbf{c}_{M \times 1}\}\mathbf{B}_{M \times \lfloor M/\alpha \rfloor}\left(\text{diag}\{\mathbf{c}_{M \times 1}\}\mathbf{B}_{M \times \lfloor M/\alpha \rfloor}\right)^H\right)^{-1}. \quad (17)$$

$()^H$ represents the conjugate transpose operation of a matrix, and $()^{-1}$ represents the inverse operation of a matrix. ∎

***Remark 3***: It is worth pointing out that, the condition in (14) can be satisfied under all the six waveforms shown in Table I. Therefore, by optimizing both the amplitude and phase of sequence elements, all the six waveforms can achieve the same optimal Pareto union bound on PAPR and AFs, as demonstrated by the numerical results in Section V.

***Remark 4***: In practical systems, sequence configuration and storage overhead need to be considered, and thus it is necessary to have constraints on the amplitude or phase of sequence elements. For example, unimodular sequences with discrete phase are preferred. In addition to overhead reduction, unimodular sequences are also beneficial for good channel estimation performance when the same sequences are reused for communication scenarios. Under these constraints on sequence elements, different waveforms will exhibit different PAPR and AF performance. Therefore, both the waveform parameters and sequences should be designed to achieve the optimal tradeoff between PAPR and AFs, and more insights will be provided in Section V.

## IV. THE PROPOSED FRAMEWORK FOR OPTIMIZING WAVEFORM PARAMETERS AND SEQUENCES

To achieve the optimal Pareto union bound on PAPR and AFs with reasonable computational complexity, we propose a framework to optimize waveform parameters and sequences jointly. In this framework, a multi-objective optimization problem under fractional delay and fractional Doppler shift is firstly formulated, and then AWSGNet is proposed to solve this problem. As mentioned in Remark 3 and Remark 4, sequences without any constraints have the potential to achieve the optimal Pareto union bound, while unimodular sequences with discrete phase are preferred in terms of realistic engineering aspects (e.g., storage overhead) and channel estimation performance. The proposed framework is capable of adapting to different types of sequences, e.g., sequences with continuous amplitude and continuous phase, unimodular sequences with continuous phase, as well as unimodular sequences with discrete phase.

### A. Problem Formulation

The multi-objective optimization problem can be formulated as

$$\min_{\mathbf{V}} \omega_1 APSL_{\mathbf{U}} + \omega_2 CPSL_{\mathbf{U}}$$
$$\text{s.t. } PAPR_{\mathbf{U}} \leq p_{th}, \quad (18)$$

where $\mathbf{V} = \{\mathbf{A}_{\lfloor K/\beta \rfloor \times K}, \mathbf{B}_{M \times \lfloor M/\alpha \rfloor}, \mathbf{c}_{M \times 1}, \mathbf{U}\}$ and $p_{th}$ is the maximum PAPR threshold. $\omega_1$ and $\omega_2$ represent the weight coefficients corresponding to APSL and CPSL, respectively, where $\omega_1, \omega_2 \in [0, 1]$ and $\omega_1 + \omega_2 = 1$. Note that different values of $\omega_1, \omega_2$ and $p_{th}$ can be set in different scenarios. For example, CPSL is not an important metric if there is only one transmitter on the same time-frequency resources, and thus $\omega_2$ can be set to 0. Furthermore, the constrained optimization problem in (18) can be converted to an unconstrained optimization problem based on the $l_1$ penalty function method [23], which can be expressed as

$$\min_{\mathbf{V}} \omega_1 APSL_{\mathbf{U}} + \omega_2 CPSL_{\mathbf{U}} + \sigma \max(PAPR_{\mathbf{U}} - p_{th}, 0), (19)$$

where the last term represents the $l_1$ penalty function and $\sigma > 0$ is the penalty factor.

To solve the multi-objective optimization problem in (19), gradient descent methods can be used to iteratively update the optimization variables. However, it is very complicated to find the optimal values of all variables in $\mathbf{V}$ simultaneously. To simplify the optimization, some variables can be set to specific values to optimize other variables, and then different values of these variables may be traversed. For example, the sequence group set $\mathbf{U}$ and the FDSS vector $\mathbf{c}_{M \times 1}$ can be optimized under one preprocessing matrix in set $\mathbb{A}$ and one preprocessing matrix in set $\mathbb{B}$, where the sets $\mathbb{A}$ and $\mathbb{B}$ include a plurality of candidate values of $\mathbf{A}_{\lfloor K/\beta \rfloor \times K}$ and $\mathbf{B}_{M \times \lfloor M/\alpha \rfloor}$, respectively. This process can be repeated to find the optimal solution by traversing different values in the sets $\mathbb{A}$ and $\mathbb{B}$. Moreover, inspired by the model training methods based on gradient descent in the field of machine learning, some powerful optimizers such as the Adam optimizer and the RMSProp optimizer [24] [25] can be utilized to efficiently calculate the gradients and update the optimization variables. To this end, AWSGNet is proposed in the next subsection to solve the optimization problem in (19).

### B. Adaptive Waveform and Sequence Generation Network

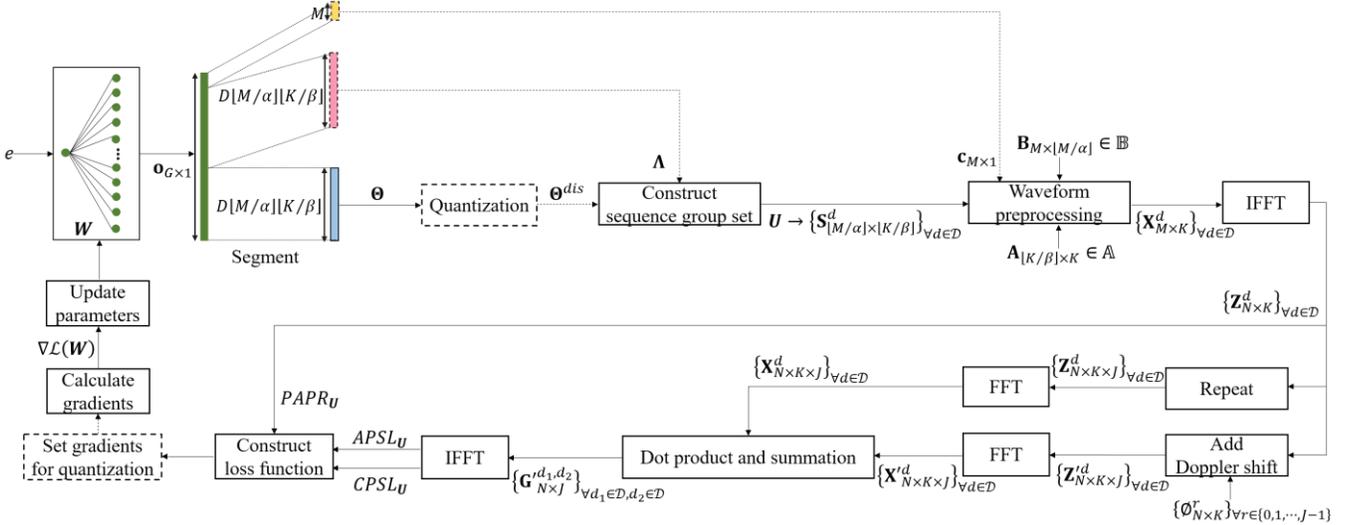

**Fig. 3** The proposed AWSGNet.

In the proposed AWSGNet shown in Fig. 3, the optimization variables in (19) can be obtained according to the output of a fully connected network, and the corresponding loss function is constructed based on the optimization variables for calculating the gradients and updating the network parameters. Specifically, only one fully connected layer is included in the fully connected network, where the input layer and the output layer consist of one neuron and $G$ neurons, respectively. The optimization can be implemented by iteratively performing forward propagation and backward propagation.

In the forward propagation, the sequence group set $U$ is generated according to the output of the fully connected network, and then the delay-time domain signal set $\{\mathbf{Z}_{N\times K}^d\}_{\forall d\in\mathcal{D}}$ can be obtained based on $U$ and waveform parameters. The output of the fully connected network can be expressed as $\mathbf{o}_{G\times 1} = \mathbf{W} \cdot e$, where $e \neq 0$ is the input of the fully connected network and $\mathbf{W}$ is a $G \times 1$-dimensional vector consisting of $G$ network parameters. Note that the value of $G$ depends on the number of optimization variables, which will be described in detail later. Based on the network output $\mathbf{o}_{G\times 1}$, the sequence group set $U$ can be constructed, or alternatively, both the sequence group set $U$ and the FDSS vector $\mathbf{c}_{M\times 1}$ can be constructed. Then, by selecting a preprocessing matrix in the set $\mathbb{A}$ and a preprocessing matrix in the set $\mathbb{B}$, the time-frequency domain signal set $\{\mathbf{X}_{M\times K}^d\}_{\forall d\in\mathcal{D}}$ is generated according to (1). Note that the sets $\mathbb{A}$ and $\mathbb{B}$ may include waveform preprocessing matrices corresponding to different values of $\alpha$ and $\beta$. Finally, the time-frequency domain signal set $\{\mathbf{X}_{M\times K}^d\}_{\forall d\in\mathcal{D}}$ can be converted to the delay-time domain signal set $\{\mathbf{Z}_{N\times K}^d\}_{\forall d\in\mathcal{D}}$ after performing $N$-point IFFT per OFDM symbol on each time-frequency domain signal in $\{\mathbf{X}_{M\times K}^d\}_{\forall d\in\mathcal{D}}$. Under different constraints of sequence elements, more details are provided as follows:

- *Sequences with continuous amplitude and continuous phase:* The sequence group set $U$ consists of $D\lfloor M/\alpha \rfloor \lfloor K/\beta \rfloor$ sequence elements, which can be generated based on the amplitude vector $\mathbf{\Lambda}$ and the phase vector $\mathbf{\Theta}$. Both $\mathbf{\Lambda}$ and $\mathbf{\Theta}$ are $D\lfloor M/\alpha \rfloor \lfloor K/\beta \rfloor \times 1$-dimensional vectors, corresponding to $D\lfloor M/\alpha \rfloor \lfloor K/\beta \rfloor$ amplitudes and $D\lfloor M/\alpha \rfloor \lfloor K/\beta \rfloor$ phases, respectively. For example, a sequence element $s$ can be represented by $s = ae^{j\theta}$, where the amplitude $a$ of $s$ is an element of $\mathbf{\Lambda}$, and the phase $\theta$ of $s$ is an element of $\mathbf{\Theta}$. If both the sequence group set $U$ and the FDSS vector $\mathbf{c}_{M\times 1}$ are optimized, we have $G = M + 2D\lfloor M/\alpha \rfloor \lfloor K/\beta \rfloor$. $M$ elements in $\mathbf{o}_{G\times 1}$ correspond to the FDSS vector $\mathbf{c}_{M\times 1}$, and other $2D\lfloor M/\alpha \rfloor \lfloor K/\beta \rfloor$ elements in $\mathbf{o}_{G\times 1}$ correspond to $\mathbf{\Lambda}$ and $\mathbf{\Theta}$. If the FDSS vector $\mathbf{c}_{M\times 1}$ does not need to be optimized (e.g., diag$\{\mathbf{c}_{M\times 1}\} = \mathbf{I}_M$ or $\mathbf{c}_{M\times 1}$ is generated based on the RRC filter), we have $G = 2D\lfloor M/\alpha \rfloor \lfloor K/\beta \rfloor$.

- *Unimodular sequences with continuous phase:* In this case, all sequences in the sequence group set $U$ have the same amplitude. Only the phase vector $\mathbf{\Theta}$ is optimized to generate $U$, and the amplitudes of all sequences can be set to 1 without loss of generality. If the FDSS vector $\mathbf{c}_{M\times 1}$ is optimized, we have $G = M + D\lfloor M/\alpha \rfloor \lfloor K/\beta \rfloor$, otherwise we have $G = D\lfloor M/\alpha \rfloor \lfloor K/\beta \rfloor$.

- *Unimodular sequences with discrete phase:* The discrete phase vector $\mathbf{\Theta}^{dis}$ can be obtained by performing a quantization operation on each element in $\mathbf{\Theta}$, i.e., $\mathbf{\Theta}^{dis} \triangleq q_t(\mathbf{\Theta})$. Specifically, the phase values of all sequence elements belong to a finite set $\Omega$, and the number of candidate phases in $\Omega$ is $B$. For example, $\Omega = \{\frac{\pi}{4}, \frac{3\pi}{4}, \frac{5\pi}{4}, \frac{7\pi}{4}\}$ if $B = 4$. For this quantization operation, the principle of proximity can be applied, i.e., a continuous phase belonging to $[0, 2\pi)$ is quantized into the discrete phase with the smallest absolute difference from it in $\Omega$, which can be expressed by

$$\theta_i^{dis} = \underset{\theta_q \in \Omega}{\arg\min} |\theta_i \bmod(2\pi) - \theta_q|, \quad (20)$$

where $\theta_i$ is the $i$th element of $\mathbf{\Theta}$ and $\theta_i^{dis}$ is the $i$th element of $\mathbf{\Theta}^{dis}$. Note that the value of $\theta_i$ may exceed $2\pi$, thus the modulo operation is performed on $\theta_i$ before the quantization operation.

During the backward propagation, the network parameters are updated by minimizing the loss function. Based on the optimization problem in (19), the loss function can be represented by

$$\mathcal{L}(\mathbf{W}) = \omega_1 APSL_U + \omega_2 CPSL_U + \sigma \max(PAPR_U - p_{th}, 0). \quad (21)$$

Note that (10) is equivalent to calculating the cyclic correlation of $\mathbf{z}'^{\mathbf{P}}_k$ and $\mathbf{z}^{\mathbf{Q}}_k$, where the $n$th element of $\mathbf{z}'^{\mathbf{P}}_k = [z'^{\mathbf{P}}_{0,k}, z'^{\mathbf{P}}_{1,k}, \cdots, z'^{\mathbf{P}}_{N-1,k}]^T$ can be expressed as $z'^{\mathbf{P}}_{n,k} = z^{\mathbf{P}}_{n,k} e^{j2\pi f_d n T_s}$. The relationship between the cyclic correlation operation of two signals and the dot product operation of the DFT of one signal and the conjugation of the DFT of another signal [26] is utilized to simplify the calculation of AF in the loss function. Then we have $\mathbf{z}'^{\mathbf{P}}_k \circledast \mathbf{z}^{\mathbf{Q}}_k = \mathbf{F}_N^H\left((\mathbf{F}_N \mathbf{z}'^{\mathbf{P}}_k) \odot (\mathbf{F}_N \mathbf{z}^{\mathbf{Q}}_k)^*\right)$, where $\mathbf{F}_N$ is DFT matrix of size $N \times N$, "$\circledast$" represents the cyclic correlation operation, and "$\odot$" represents the dot product operation. Let $\mathbf{x}'^{\mathbf{P}}_k = \mathbf{F}_N \mathbf{z}'^{\mathbf{P}}_k$ and $\mathbf{x}^{\mathbf{Q}}_k = \mathbf{F}_N \mathbf{z}^{\mathbf{Q}}_k$, (10) can be rewritten as

$$AF_{\mathbf{z}^{\mathbf{P}}_k, \mathbf{z}^{\mathbf{Q}}_k}(\tau, f_d) = \frac{1}{\sqrt{N}} \sum_{n=0}^{N-1} x'^{\mathbf{P}}_{n,k} (x^{\mathbf{Q}}_{n,k})^* e^{-j\frac{2\pi n \tau}{N}}, \quad (22)$$

where $x'^{\mathbf{P}}_{n,k}$ is the $n$th element of $\mathbf{x}'^{\mathbf{P}}_k = [x'^{\mathbf{P}}_{0,k}, x'^{\mathbf{P}}_{1,k}, \cdots, x'^{\mathbf{P}}_{N-1,k}]^T$ and $x^{\mathbf{Q}}_{n,k}$ is the $n$th element of $\mathbf{x}^{\mathbf{Q}}_k = [x^{\mathbf{Q}}_{0,k}, x^{\mathbf{Q}}_{1,k}, \cdots, x^{\mathbf{Q}}_{N-1,k}]^T$. In this way, the FFT operation can be used to accelerate the calculation of AF. More specifically, the calculation process of the loss function shown in Fig. 3 is summarized as follows:

- The time-frequency domain signal set $\{\mathbf{X}^d_{N \times K \times J}\}_{\forall d \in \mathcal{D}}$ is generated by performing $N$-point FFT on each delay-time domain signal in $\{\mathbf{Z}^d_{N \times K \times J}\}_{\forall d \in \mathcal{D}}$ along the first dimension, where $\mathbf{Z}^d_{N \times K \times J}$ can be obtained by repeating the delay-time domain signal $\mathbf{Z}^d_{N \times K}$ for $J$ times, i.e., $\mathbf{Z}^d_{N \times K \times J}[:,:,r] = \mathbf{Z}^d_{N \times K}$, where $r \in \{0, 1, \cdots, J-1\}$.

- The time-frequency domain signal set $\{\mathbf{X}'^d_{N \times K \times J}\}_{\forall d \in \mathcal{D}}$ is generated by performing $N$-point FFT on each delay-time domain signal in $\{\mathbf{Z}'^d_{N \times K \times J}\}_{\forall d \in \mathcal{D}}$ along the first dimension. Specifically, $\mathbf{Z}'^d_{N \times K \times J}$ can be obtained based on $\mathbf{Z}''^{d,r}_{N \times K}$, i.e., $\mathbf{Z}'^d_{N \times K \times J}[:,:,r] = \mathbf{Z}''^{d,r}_{N \times K}$, where $r \in \{0, 1, \cdots, J-1\}$ and $\mathbf{Z}''^{d,r}_{N \times K}$ is calculated by

$$\mathbf{Z}''^{d,r}_{N \times K} = \mathbf{\Phi}^r_{N \times K} \cdot \mathbf{Z}^d_{N \times K}. \quad (23)$$

$\mathbf{\Phi}^r_{N \times K}$ is the phase shift matrix corresponding to the Doppler shift $r\Delta f_{ds}$, and the element in the $n$th row and the $k$th column of $\mathbf{\Phi}^r_{N \times K}$ is $e^{j\left((k-1)\varphi^r_s + (n-1)\varphi^r_f\right)}$, where $\varphi^r_f = 2\pi r \Delta f_{ds} T_s$, $\varphi^r_s = 2\pi r \Delta f_{ds} T_c$, $n \in \{0, 1, \cdots, N-1\}$ and $k \in \{0, 1, \cdots, K-1\}$.

- Then the set $\{\mathbf{G}'^{d_1, d_2}_{N \times J}\}_{\forall d_1 \in \mathcal{D}, d_2 \in \mathcal{D}}$ for AF calculation is generated according to $\{\mathbf{X}^d_{N \times K \times J}\}_{\forall d \in \mathcal{D}}$ and $\{\mathbf{X}'^d_{N \times K \times J}\}_{\forall d \in \mathcal{D}}$, where $\mathbf{G}'^{d_1, d_2}_{N \times J}$ can be obtained by performing the summation operation on $\mathbf{G}^{d_1, d_2}_{N \times K \times J}$ along the third dimension and $\mathbf{G}^{d_1, d_2}_{N \times K \times J}$ is expressed as

$$\mathbf{G}^{d_1, d_2}_{N \times K \times J} = \mathbf{X}'^{d_1}_{N \times K \times J} \odot \left(\mathbf{X}^{d_2}_{N \times K \times J}\right)^*. \quad (24)$$

- The AF set $\{\mathbf{G}''^{d_1, d_2}_{N \times J}\}_{\forall d_1 \in \mathcal{D}, d_2 \in \mathcal{D}}$ is generated by performing $N$-point IFFT on each signal in $\{\mathbf{G}'^{d_1, d_2}_{N \times J}\}_{\forall d_1 \in \mathcal{D}, d_2 \in \mathcal{D}}$ along the first dimension. If $d_1 = d_2$, $\mathbf{G}'^{d_1, d_2}_{N \times J}$ corresponds to an auto-AF, otherwise $\mathbf{G}'^{d_1, d_2}_{N \times J}$ corresponds to a cross-AF as defined in (9).

- Finally, the loss function in (20) can be obtained based on $APSL_U$ $CPSL_U$ and $PAPR_U$, where $APSL_U$ and $CPSL_U$ are calculated according to (12) and (13) based on the AF set $\{\mathbf{G}''^{d_1, d_2}_{N \times J}\}_{\forall d_1 \in \mathcal{D}, d_2 \in \mathcal{D}}$, and $PAPR_U$ is calculated according to (7) based on the delay-time domain signal set $\{\mathbf{Z}^d_{N \times K}\}_{\forall d \in \mathcal{D}}$.

Furthermore, the network parameters are updated using gradient descent, where the gradients of the loss function with respect to $\mathbf{W}$ should be calculated. Specifically, in the $t$th iteration, the gradient vector $\mathbf{g}_t$ is calculated by

$$\mathbf{g}_t = \nabla \mathcal{L}(\mathbf{W}_t). \quad (25)$$

Then, the network parameters can be updated by

$$\mathbf{W}_{t+1} = \mathbf{W}_t - \eta_t \mathbf{g}_t, \quad (26)$$

where $\mathbf{W}_t$ and $\mathbf{W}_{t+1}$ respectively represent the network parameters in the $t$th iteration and the $(t+1)$th iteration, and $\eta_t$ is the learning rate in the $t$th iteration. Some efforts have been done to improve convergence speed of gradient descent methods. For example, in the Adam optimizer [27], (26) can be replaced by

$$\mathbf{W}_{t+1} = \mathbf{W}_t - \frac{\eta_t}{\sqrt{\widehat{\mathbf{m}}_{2,t}} + \epsilon} \widehat{\mathbf{m}}_{1,t}, \quad (27)$$

where

$$\begin{aligned}
\mathbf{m}_{1,t} &= \rho_1 \mathbf{m}_{1,t-1} + (1 - \rho_1) \mathbf{g}_t, \\
\mathbf{m}_{2,t} &= \rho_2 \mathbf{m}_{2,t-1} + (1 - \rho_2) \mathbf{g}_t^2, \\
\widehat{\mathbf{m}}_{1,t} &= \frac{\mathbf{m}_{1,t}}{1 - (\rho_1)^t}, \\
\widehat{\mathbf{m}}_{2,t} &= \frac{\mathbf{m}_{2,t}}{1 - (\rho_2)^t}.
\end{aligned} \quad (28)$$

In the first two formulas in (28), the biased first moment estimate and the biased second raw moment estimate are updated, respectively. In the last two formulas in (28), the bias-corrected first moment estimate and the bias-corrected second moment estimate are computed. $\rho_1$ and $\rho_2$ respectively represent the exponential decay rates for moment estimates. $\epsilon$ is a small constant used for numerical stability.

It is worth noting that if the sequences are optimized under discrete phase constraints, the loss function is nondifferentiable owning to the quantization operation, which means the gradients with respect to $\mathbf{\Theta}$ cannot be directly calculated. To solve this problem, the idea of straight through estimator (STE) is adopted to avoid the gradient vanishing problem [28]. Specifically, for ease of understanding, it is assumed that $e = 1$ and only the phase vector $\mathbf{\Theta}$ is optimized, then we have $\mathbf{W} = \mathbf{\Theta}$ and the loss function in (21) can be rewritten as

**Algorithm 1** Main steps of the proposed AWSGNet.

**Input:** $\mathbb{A}$, $\mathbb{B}$, $\omega_1$, $\omega_2$, $p_{th}$, $\sigma$, $f_D$, $J$, $b$, $N$, $T_s$, $T_c$, $B$, $e$, initial network parameters $\mathbf{W}_{initial}$, and the number of iterations $T$.

**Output:** Optimized FDSS vectors and sequence group sets under different preprocessing matrices.

1. **For** $\mathbf{A}_{\lfloor K/\beta \rfloor \times K} \in \mathbb{A}$ and $\mathbf{B}_{M \times \lfloor M/\alpha \rfloor} \in \mathbb{B}$
2. Set $t = 0$.
3. Initialize $\mathbf{W}_0 = \mathbf{W}_{initial}$.
4. Obtain $\mathbf{c}_{M \times 1}$, $\mathbf{\Lambda}$ and $\mathbf{\Theta}$ based on $\mathbf{W}_t$.
5. **If** sequences with discrete phase are considered
6. Obtain $\mathbf{\Theta}^{dis}$ according to (20).
7. **End If**
8. Construct the sequence group set $\mathbf{U}$.
9. Obtain $\{\mathbf{X}_{M \times K}^d\}_{\forall d \in \mathcal{D}}$ based on $\mathbf{U}$, $\mathbf{A}_{\lfloor K/\beta \rfloor \times K}$, $\mathbf{B}_{M \times \lfloor M/\alpha \rfloor}$, and $\mathbf{c}_{M \times 1}$.
10. Obtain $\{\mathbf{Z}_{N \times K}^d\}_{\forall d \in \mathcal{D}}$ based on $\{\mathbf{X}_{M \times K}^d\}_{\forall d \in \mathcal{D}}$.
11. Obtain $\{\mathbf{X}'^{d}_{N \times K \times J}\}_{\forall d \in \mathcal{D}}$ and $\{\mathbf{X}^d_{N \times K \times J}\}_{\forall d \in \mathcal{D}}$ based on $\{\mathbf{Z}_{N \times K}^d\}_{\forall d \in \mathcal{D}}$.
12. Obtain $\{\mathbf{G}'^{d_1,d_2}_{N \times J}\}_{\forall d_1 \in \mathcal{D}, d_2 \in \mathcal{D}}$ based on $\{\mathbf{X}'^{d}_{N \times K \times J}\}_{\forall d \in \mathcal{D}}$ and $\{\mathbf{X}^d_{N \times K \times J}\}_{\forall d \in \mathcal{D}}$.
13. Obtain $\{\mathbf{G}''^{d_1,d_2}_{N \times J}\}_{\forall d_1 \in \mathcal{D}, d_2 \in \mathcal{D}}$ based on $\{\mathbf{G}'^{d_1,d_2}_{N \times J}\}_{\forall d_1 \in \mathcal{D}, d_2 \in \mathcal{D}}$.
14. Calculate $APSL_U$ and $CPSL_U$ based on $\{\mathbf{G}''^{d_1,d_2}_{N \times J}\}_{\forall d_1 \in \mathcal{D}, d_2 \in \mathcal{D}}$.
15. Calculate $PAPR_U$ based on $\{\mathbf{Z}_{N \times K}^d\}_{\forall d \in \mathcal{D}}$.
16. Calculate the value of the loss function in (21) or (29).
17. **If** sequences with discrete phase are considered
18. Set $\nabla q_i(\mathbf{\Theta}_t) = 1$.
19. **End If**
20. Calculate the gradients according to (25) or (30).
21. Update $\mathbf{W}_{t+1}$ according to (27) and (28).
22. **Stopping Criterion**: if $t = T$, stop, else update $t = t + 1$ and turn to Step 4.
23. **End for**

$$\mathcal{L}(\mathbf{\Theta}^{dis}) = \mathcal{L}(q_i(\mathbf{\Theta})) = \omega_1 APSL_U + \omega_2 CPSL_U + \sigma \max(PAPR_U - p_{th}, 0). \quad (29)$$

Then, the gradient vector $\mathbf{g}'_t$ in the $t$ th iteration can be calculated by

$$\mathbf{g}'_t = \nabla \mathcal{L}(q_i(\mathbf{\Theta}_t)) \cdot \nabla q_i(\mathbf{\Theta}_t), \quad (30)$$

where $\nabla q_i(\mathbf{\Theta}_t)$ represents the gradient of the quantization function $q_i()$ with respect to $\mathbf{\Theta}_t$. Since $\nabla q_i(\mathbf{\Theta}_t)$ is nondifferentiable as mentioned before, the value of $\nabla q_i(\mathbf{\Theta}_t)$ can be set to 1, and thus $\mathbf{g}'_t = \nabla \mathcal{L}(q_i(\mathbf{\Theta}_t))$. Particularly, the gradient vector $\mathbf{g}_t$ used in (26) and (28) can be replaced with $\mathbf{g}'_t$ in this case.

Taking the case where the FDSS vector is optimized as an example, the main steps of the proposed AWSGNet are summarized in Algorithm 1. Note that the preprocessing matrix sets $\mathbb{A}$ and $\mathbb{B}$ can be regarded as the input of Algorithm 1. By traversing different preprocessing matrices in $\mathbb{A}$ and $\mathbb{B}$, the optimal sequences under different waveforms can be obtained. Besides, the parameter $b$ in (11) is also one of the input parameters of Algorithm 1, which is used to restrict the main lobe width of auto-AF. Moreover, the input $\mathbf{W}_{initial}$ of Algorithm 1 can also be decomposed into $\mathbf{c}_{M \times 1}^{initial}$, $\mathbf{\Lambda}_{initial}$ and $\mathbf{\Theta}_{initial}$. For example, $\mathbf{c}_{M \times 1}^{initial}$ and $\mathbf{\Lambda}_{initial}$ can be obtained by setting all elements to 1, $\mathbf{\Theta}_{initial}$ can be obtained by setting each element to be randomly distributed among $[0, 2\pi)$. Alternatively, $\mathbf{\Theta}_{initial}$ can also be obtained based on phases of the existing sequences such as ZC and Gold sequences.

V. NUMERICAL RESULTS

In this section, some practical design examples based on the proposed framework in Section IV are provided. Moreover, under different sequence element constraints, the performance of the optimized waveforms and sequences is compared with that of the existing 5G waveforms and sequences in terms of PAPR, APSL and CPSL.

*A. Setup*

Unless otherwise noted, the parameters described in this subsection are used for performance evaluation. Specifically, we consider that $D = 30$, $K = 4$, $M = 204$, $N = 2048$, $\Delta f = 120$ kHz, and $T_c$ equals to the duration of one OFDM symbol. In addition, it is assumed that $\omega_1 = \omega_2 = 0.5$, $p_{th} = 2$ dB, $\sigma = 1$, $f_D = 7.2$ kHz, $J = 9$, $T = 5000$, $e = 1$ and the Adam optimizer with $\rho_1 = 0.9$, $\rho_2 = 0.999$ and $\epsilon = 10^{-8}$ is used in the proposed AWSGNet.

The baseline schemes for comparison include "CP-OFDM + Gold", "CP-OFDM + ZC", "DFT-s-OFDM + Gold" and "DFT-s-OFDM + Gold + FDSS", as explained in Section III. Based on the proposed AWSGNet, the optimized sequences under all the six waveforms shown in Table I can be obtained, where $W_1 \approx 1$ if FDSS is not used, and the value of $W_1$ will be described later if FDSS is adopted. In FTN-s-OFDM and FTN-s-OTFS waveforms, $\alpha = 0.5$ is assumed. In DFTN-s-OTFS waveform, we set $\alpha = 0.5$ and $\beta = 0.5$. "*WaveformX* + proposed" and "*WaveformX* + proposed + FDSS" indicate the performance of the optimized sequences under "*WaveformX*" with and without FDSS, respectively, where "*WaveformX*" can be replaced by any waveform in Table I.

*B. Performance for Sequences with Continuous Amplitude and Continuous Phase*

We firstly evaluate the performance of the designed sequences with continuous amplitude and continuous phase. The performance comparison in terms of APSL, CPSL and PAPR is shown in Fig. 4. In this case, some observations are highlighted as follows:

- The optimal Pareto union bound shown in Fig. 2 can be achieved by the designed sequences under all the six waveforms as revealed in Theorem 1. Specifically, the values of APSL and CPSL under all these designed mechanisms are -20.6 dB and -24.2 dB, respectively, and the maximum PAPR is 2 dB. This result is consistent with the performance corresponding to one point on the APSL-CPSL curve in Fig. 2 (b) at a maximum PAPR of

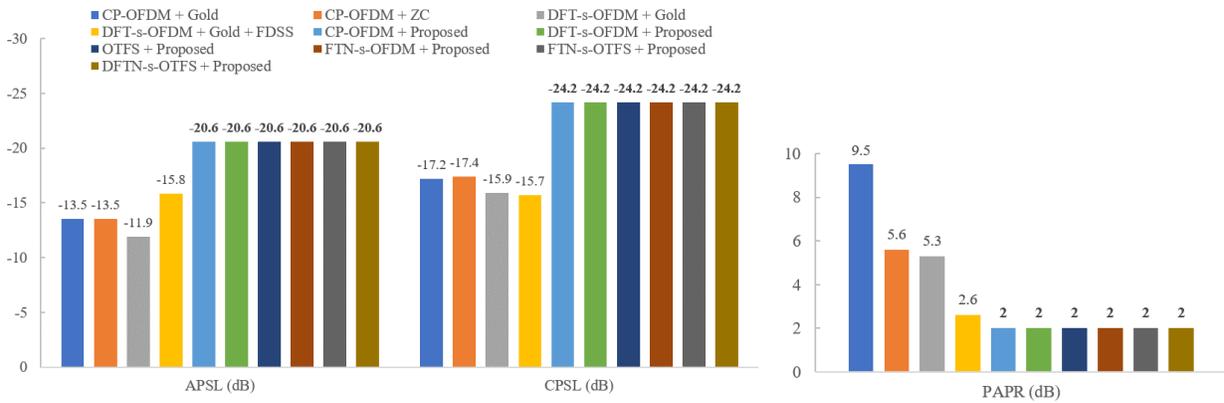

**Fig. 4.** Performance of the designed sequences with continuous amplitude and continuous phase.

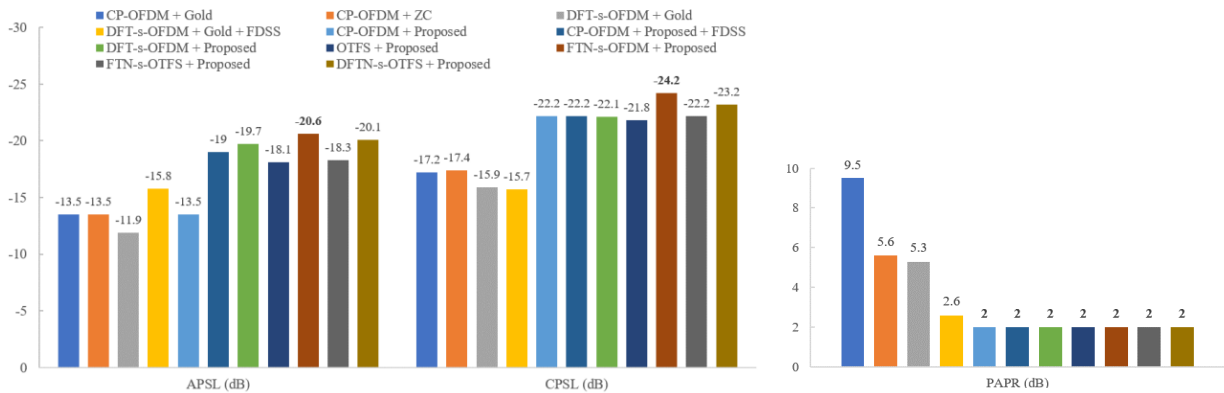

**Fig. 5.** Performance of the designed unimodular sequences with continuous phase.

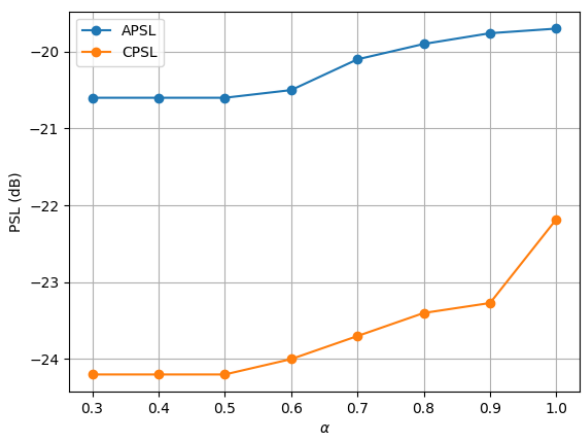

**Fig. 6.** APSL and CPSL performance of the designed unimodular sequences with continuous phase under FTN-s-OFDM waveform using different values of $\alpha$.

2 dB. Note that the same performance can be observed with and without FDSS in the proposed schemes, and thus only the performance without FDSS is shown in Fig. 4.

- Compared to the state-of-the-art 5G waveforms and sequences, the designed schemes can achieve significant improvements in APSL, CPSL, and PAPR performance. For example, compared to CP-OFDM waveform using ZC sequences, the APSL gain is 7.1 dB, the CPSL gain is 6.8 dB, and the PAPR gain is 3.6 dB.

*C. Performance for Unimodular Sequences with Continuous Phase*

We further evaluate the performance of the designed unimodular sequences with continuous phase, where the maximum PAPR is 2 dB. Fig. 5 shows the performance comparison in terms of APSL, CPSL and PAPR. In addition, the APSL and CPSL performance of the designed unimodular sequences with continuous phase under FTN-s-OFDM waveform using different values of $\alpha$ is shown in Fig. 6. From Fig. 5 and Fig. 6, the following observations can be summarized.

- As shown in Fig. 5, different waveforms exhibit different performance if only the phases of the sequence elements are optimized. It is quite surprising that the optimal Pareto union bound in Fig. 2 can still be achieved by FTN-s-OFDM waveform. Therefore, in scenarios where PAPR, APSL and CPSL are all important, FTN-s-OFDM waveform may be preferred. In addition, compared with OTFS, FTN-s-OTFS and DFTN-s-OTFS can achieve better APSL and CPSL performance under low PAPR constraints, thanks to the integration of FTN.
- The designed unimodular sequences with continuous phase still perform better than the existing sequences under any of the waveforms in Table I. For example, under DFT-s-OFDM waveform, the proposed sequences can achieve an APSL gain of 7.8 dB, a CPSL gain of 6.2 dB and a PAPR gain of 3.3 dB compared to Gold sequences.

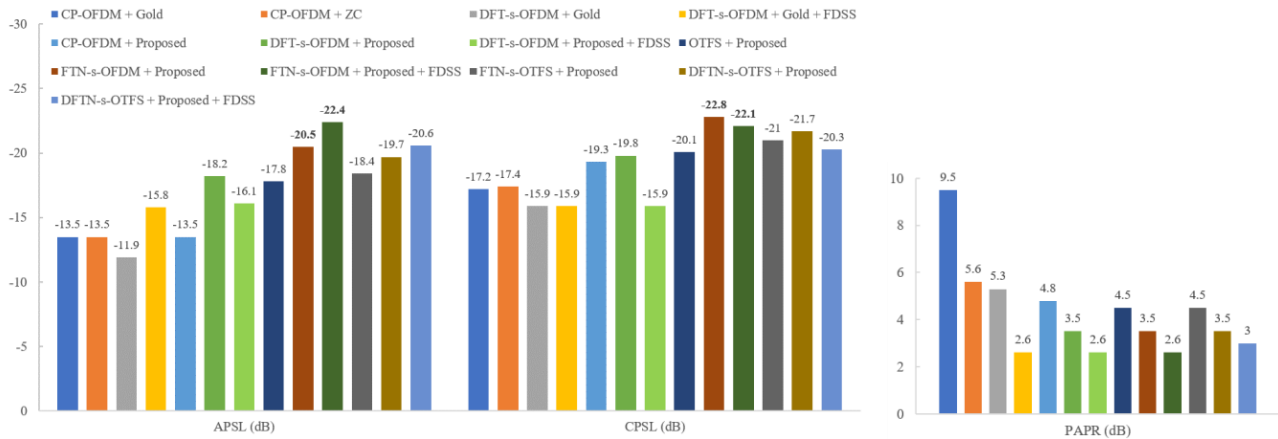

**Fig. 7.** Performance of the designed unimodular sequences with discrete phase.

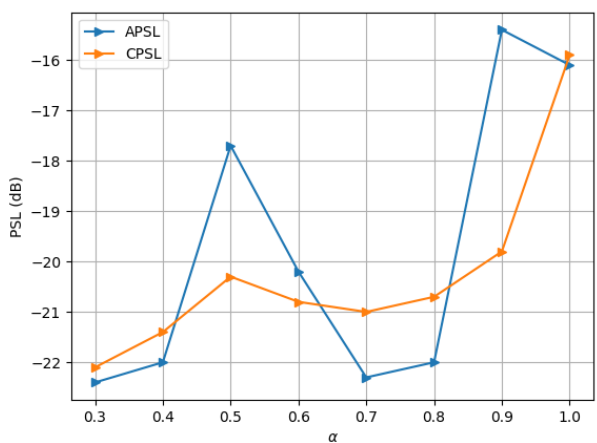

**Fig. 8.** APSL and CPSL performance of the designed unimodular sequences with discrete phase under FTN-s-OFDM waveform using different values of $\alpha$.

- The performance comparison with and without FDSS under CP-OFDM waveform is also shown in Fig. 5. As mentioned in Section II-A, the auto-AF is a two-dimensional sinc-like function if FDSS is not adopted, and thus the same APSL performance can be observed by using the designed sequences and the existing 5G sequences. However, an APSL gain of 5.5 dB can be achieved by optimizing the FDSS vector without degrading the PAPR and CPSL performance, where $W_1 \approx 1$. Note that the same performance can be observed with and without FDSS when the sequences are optimized under other waveforms in Table I.
- From Fig. 6, it can be found that the smaller the value of $\alpha$, the better the performance of APSL and CPSL. Nevertheless, the performance tends to converge to the optimal values when $\alpha \leq 0.5$. Therefore, setting $\alpha$ to a value around 0.5 may be a wise choice to achieve a good tradeoff between performance and overhead.

### D. Performance for Unimodular Sequences with Discrete Phase

The performance of the designed unimodular sequences with discrete phase is evaluated in this subsection, where $B = 4$ is assumed. In this case, the following conclusions can be obtained.

- The APSL, CPSL and PAPR performance of different schemes is shown in Fig. 7. Similar to the observation from Fig. 5, FTN-s-OFDM waveform with $\alpha = 0.3$ can still perform better than the other waveforms even under the constraints of discrete phase of sequence elements. In addition, some performance degradation can be observed compared to the case where unimodular sequences with continuous phase are optimized. For example, under FTN-s-OFDM waveform without FDSS, the performance of APSL, CPSL and PAPR using unimodular sequences with discrete phase is degraded by 0.1 dB, 2.4 dB, and 1.5 dB respectively compared to the performance using unimodular sequences with continuous phase.
- Nevertheless, the designed unimodular sequences with discrete phase perform better than the existing sequences under any of the waveforms in Table I. For example, under DFT-s-OFDM waveform without FDSS, the proposed sequences can achieve an APSL gain of 6.3 dB, a CPSL gain of 3.9 dB and a PAPR gain of 1.8 dB compared to Gold sequences.
- It can be observed from Fig. 7 that FDSS is beneficial for PAPR reduction, while good APSL and CPSL performance can still be maintained. For example, under FTN-s-OFDM waveform using the optimized FDSS vector, the maximum PAPR is 2.6 dB, APSL is -22.4 dB and CPSL is -22.1 dB. Note that $W_1 \approx 1.4$ when FDSS is used in the proposed schemes.
- Moreover, Fig. 8 shows the APSL and CPSL performance of the designed unimodular sequences with discrete phase under FTN-s-OFDM waveform using different values of $\alpha$, where FDSS is also optimized and the maximum PAPR is 2.6 dB. Different from Fig. 6, the conclusion that the smaller the value of $\alpha$, the better the performance is not always true. Therefore, the value of $\alpha$ should be carefully selected to achieve a good sensing performance.

## VI. CONCLUSION

In this paper, the optimal tradeoff between PAPR, auto-AF, and cross-AF has been investigated to achieve both good coverage and interference rejection capability in ISAC systems. Firstly, a generalized OFDM waveform set with a unified parametric representation was proposed, which include the existing 5G waveforms and some new waveforms. Then, the optimal Pareto union bound on PAPR and AFs was developed for the generalized OFDM waveform set, which can provide guidance for the design of waveform parameters and sequences. To achieve the optimal Pareto union bound with reasonable computational complexity, a framework was established to jointly optimize waveform parameters and sequences, where AWSGNet was proposed to solve the multi-objective optimization problem under fractional delay and fractional Doppler shift. The results show that the designed waveforms and sequences can achieve lower PAPR, APSL and CPSL compared to the existing 5G waveforms and sequences. Particularly, to avoid the gradient vanishing problem when considering unimodular sequences with discrete phase, the idea of STE was adopted in this paper. For future research, more sophisticated optimization algorithms can be studied to improve the performance under discrete phase constraints.